\documentclass[reprint,superscriptaddress]{revtex4-1}
\usepackage{graphicx}
\usepackage{amssymb}
\usepackage{amsmath}
\usepackage[product-units = power]{siunitx}
\usepackage{amsfonts}
\usepackage{bbold}
\usepackage[version=3]{mhchem}
\usepackage[citecolor=blue,colorlinks=true]{hyperref}
\usepackage[dvipsnames]{xcolor}
\usepackage{epstopdf}

\DeclareSIUnit\BohrMagneton{\ensuremath{\mu_\textrm{B}}}
\DeclareSIUnit\formulaunit{f.u.}
\DeclareSIUnit\atomicunit{a.u.}
\DeclareSIUnit\arbunit{arb.unit}
\DeclareSIUnit\torr{Torr}
\DeclareSIUnit\counts{Counts}
\DeclareSIUnit\curie{Ci}

\usepackage{hyperref}
\hypersetup{
    bookmarks=true,         
    unicode=true,          
    pdftoolbar=true,        
    pdfmenubar=true,        
    pdffitwindow=false,     
    pdfstartview={FitB},    
    pdfborder={0 0 0},      
    pdftitle={Structure, element-specific magnetism and magneto-transport properties of
	epitaxial D0_${22}$ Mn$_2$Fe$_x$Ga thin films},
    pdfauthor={Davide Betto et.al.},     
    pdfsubject={Mn$_2$Fe$_x$Ga full characterisation},   
    pdfcreator={Davide Betto},   
	pdfkeywords={Heusler, spintronics, half metals},
    pdfnewwindow=true,      
    colorlinks=true,       
    linkcolor=red,          
    citecolor=green,        
    filecolor=magenta,      
    urlcolor=cyan,           
    linktocpage=true,       
    hyperindex=true,        
    pdfpagelabels=true,
    plainpages=false,
}

\begin{document}

\title{Structure, site-specific magnetism and magneto-transport properties of epitaxial D0$_{22}$
Mn$_2$Fe$_x$Ga thin films}

\author{Davide Betto}
\affiliation{European Synchrotron Radiation Facility, B.P. 220, 38043 Grenoble, France}
\author{Yong-Chang Lau}
\email[Electronic mail: ]{yongchang.lau@qspin.phys.s.u-tokyo.ac.jp }
\altaffiliation[Present address: ]{Department of Physics, The University of Tokyo, Bunkyo, Tokyo 113-0033, Japan}
\affiliation{CRANN and AMBER, Trinity College, Dublin 2, Ireland}
\author{Kiril Borisov}
\affiliation{CRANN and AMBER, Trinity College, Dublin 2, Ireland}
\author{Kurt Kummer}
\affiliation{European Synchrotron Radiation Facility, B.P. 220, 38043 Grenoble, France}
\author{N.B. Brookes}
\affiliation{European Synchrotron Radiation Facility, B.P. 220, 38043 Grenoble, France}
\author{Plamen Stamenov}
\affiliation{CRANN and AMBER, Trinity College, Dublin 2, Ireland}
\author{J.M.D. Coey}
\affiliation{CRANN and AMBER, Trinity College, Dublin 2, Ireland}
\author{Karsten Rode}
\affiliation{CRANN and AMBER, Trinity College, Dublin 2, Ireland}

\begin{abstract}
  Ferrimagnetic \ce{Mn2Fe_xGa} $(0.26 \leq x \leq 1.12)$ thin films have been
  characterised by X-ray diffraction, SQUID magnetometry, X-ray
  absorption spectroscopy, X-ray magnetic circular dichroism and M\"{o}ssbauer spectroscopy with the aim of
  determining the structure and site-specific magnetism
  of this tetragonal, D0$_{22}$-structure Heusler compound. High-quality epitaxial
  films with low RMS surface roughness ($\sim \SI{0.6}{\nano\metre}$) are grown by magnetron co-sputtering.
  The tetragonal distortion induces strong perpendicular magnetic anisotropy along the $c$-axis
  with a typical coercive field  $\mu_0 H\sim \SI{0.8}{\tesla}$ and an
  anisotropy field ranging from $6$ to \SI{8}{\tesla}.
  Upon increasing the Fe content $x$, substantial uniaxial anisotropy, $K_\mathrm{u} \geq
  \SI{1.0}{\mega\joule\per\metre\cubed}$ can be maintained over the full $x$ range,
  while the magnetisation of the compound is reduced from $400$ to
  $\SI{280}{\kilo\ampere\per\metre}$.
  The total magnetisation is almost entirely given by the sum of the spin moments
  originating from the ferrimagnetic Mn and Fe
  sublattices, with the latter being coupled ferromagnetically to one of the former. The orbital
  magnetic moments are practically quenched, and have negligible contributions to the
  magnetisation. The films with $x=0.73$ exhibit a high anomalous Hall angle of
  $\SI{2.5}{\percent}$ and a high Fermi-level spin polarisation, above
  $\SI{51}{\percent}$, as measured by point contact Andreev reflection.
  The Fe-substituted \ce{Mn2Ga} films are highly tunable with a unique combination of high
  anisotropy, low magnetisation, appreciable spin polarisation and low surface roughness,
  making them very strong candidates for thermally-stable spin-transfer-torque
  switching nanomagnets with lateral dimensions down to \SI{10}{\nano\metre}.
  \end{abstract}
\pacs{not.a.number}
\date{\today}

\maketitle
\section{Introduction}\label{sec:introduction}

Magnetic materials that exhibit a combination of strong uniaxial anisotropy, low magnetisation and high spin polarisation
are crucial for the development of magnetic tunnel junction (MTJ)-based spin-transfer torque (STT)
memories and oscillators.
The thermal stability of a memory cell is determined by the factor $\Delta = K_\textrm{eff} V / k_\mathrm{B}T$ where
$K_\textrm{eff}$, $V$ and $k_\mathrm{B}$
are the effective anisotropy constant, the cell volume and the Boltzmann constant, respectively.
$\Delta > 60$ is conventionally required for ten-year data retention.
A thermally-stable storage element with lateral dimensions below \SI{10}{\nano\metre} and
a thickness of less than \SI{3}{\nano\metre} therefore requires a material or structure with
$K_\textrm{eff} \sim \SI{1}{\mega\joule\per\metre\cubed}$.

Currently, the most studied storage system is Ta/ultrathin CoFeB/MgO-based heterostructures
where the perpendicular magnetic anisotropy (PMA) is obtained via
the surface/interface anisotropy of CoFeB/MgO\cite{Ikeda2010,Worledge2011,Gan2011}.
The structure exhibits a moderate $K_\textrm{eff} \sim \SI{0.2}{\mega\joule\per\metre\cubed}$,
due to the competition between interface and shape anisotropy associated with the sizable magnetisation.
$K_\textrm{eff}$ can often be improved by introducing a second CoFeB/MgO interface
in more complicated MgO/CoFeB/Ta/CoFeB/MgO structures\cite{Sato2012}.
Nevertheless, it has been shown that even
the optimised MTJ structure is unstable for dimensions below \SI{30}{\nano\metre}\cite{Sato2014}.
Novel materials with strong magnetocrystalline or strain-induced PMA are needed.

A number of Mn-based Heusler alloys crystallise in
the tetragonal D0$_{22}$ structure\cite{Graf2011,Balke2013}.
It is a variant of the cubic $L2_1$ structure with reduced symmetry, where the lattice $c$ parameter
is increased, giving rise to strong magneto-crystalline anisotropy.
The potential of this material class in STT applications was
first highlighted by Balke \textit{et al.}\cite{Balke2007}, based on the bulk properties of
D0$_{22}$ \ce{Mn3$_{x}$Ga}.
This pioneering work has led to the growth and characterisation of Mn-Ga thin films
with properties that fulfil all major requirements for STT applications,
\emph{i.e.}, strong uniaxial anisotropy\cite{Wu2009}, low magnetisation, high spin polarisation
\cite{Kurt-PRB-2011}, low damping, and high resonance
frequency\cite{Mizukami2011,Awari2016}.
In addition, the magnetic properties have been shown to be tunable by atomic substitution\cite{Winterlik2012}.
Other examples of Mn-based tetragonal Heusler compounds with high PMA are \ce{Mn3Ge}\cite{Kurt2012},
\ce{Mn$_{3-x}$Co$_x$Ga}\cite{Ouardi2012,Fowley2015} and \ce{Mn2Fe$_x$Ga}\cite{Gasi2013,Niesen2016}.
A particular case to  note is \ce{Mn2Ru$_{x}$Ga}, which has a moderate anisotropy energy ($K_\textrm{eff} \sim \SI{40}{\kilo\joule\per\metre\cubed}$),
but an extremely low magnetisation leading to a very high predicted resonance
frequency\cite{Kurt2014,Thiyagarajah2015,Betto2015,Betto2016,Zic2016,Borisov2016}.

Gasi \emph{et al.}\cite{Gasi2013} have synthesised polycrystalline ingots of \ce{Mn2FeGa}, in
both tetragonal and pseudo-cubic structures. However, the tetragonal samples
exhibit an exchange-spring behaviour with low
remanence and low magnetic anisotropy. This was attributed to the presence of
two magnetic phases, possibly due to Mn-Fe atomic disorder.
Recently, Niesen \emph{et al.}\cite{Niesen2016} deposited Mn-Fe-Ga thin films with
various compositions in both cubic and tetragonal phases. They found maximal
coercivity (\SI{1.8}{\tesla}) for a \ce{Mn3Fe_{0.4}Ga} composition.

Here we determine the structural, magnetic and magneto-transport properties
of the tetragonally distorted Heusler alloy \ce{Mn2Fe_xGa} (MFG), with $0.26\leq x\leq 1.12$,
in the form of epitaxial thin films. We used X-ray diffraction (XRD), SQUID magnetometry
and magnetotransport to characterise the overall features of the
samples, while the use of synchrotron radiation-based XAS and XMCD, as well as
M\"{o}ssbauer spectroscopy, allowed us to distinguish between the
different magnetic elements.

\section{Experimental details}\label{sec:experimental_details}
High-quality epitaxial \ce{Mn2Fe$_x$Ga} films are grown on \SI{10 x 10}{\milli\metre}
single-crystal MgO(001) substrates in an automated Shamrock-based sputtering cluster tool with a base pressure of
\SI{4E-8}{\milli\bar}. The substrate temperature is kept at $T_\textrm{sub}=\SI{300}{\degreeCelsius}$.
The \ce{Fe} concentration $x$
is varied from $x\sim0.26$ to $x\sim1.12$ by tuning the power of the \ce{Fe} gun while keeping that of the
\ce{Mn2Ga} gun fixed during co-sputtering. The growth time has been \SI{30}{\minute} for all samples, and we obtain MFG films of
thicknesses ranging from \SIrange{41.5}{50.5}{\nano\metre}, corresponding to an MFG growth rate
of about \SI{0.025}{\nano\metre\per\second}. The samples are then capped with \SI{2}{\nano\metre}
\ce{AlO$_x$} to avoid oxidation. We find, from the X-ray reflectometry, that the film thickness increases linearly as a function
of the \ce{Fe} sputtering power, while the X-ray density remains practically constant. We hence estimate the \ce{Fe} concentration in
our samples, assuming \emph{no} crystalline vacancies nor interstitials. We will further discuss
the validity of such an assumption in Section \ref{sec:structural_magnetic_prop}.
Strictly speaking, the exact formula unit of
\ce{Mn2Fe$_x$Ga} should therefore always contain 4 atoms, e.g., \ce{Mn2Fe_{0.26}Ga} should be written as \ce{Mn_{2.46}Fe_{0.31}Ga_{1.23}}.
We shall, for readability, keep the formula unit with only one variable $x$ throughout the text,
and implicitly normalise the atom count to 4 whenever necessary.

The crystal structure and lattice parameters have been determined by symmetrical $\theta - 2\theta$ scans and
reciprocal space maps using a BRUKER D8 diffractometer.
The primary beam optical path contains a Cu $K_{\alpha}$ X-ray tube with a G\"{o}bel mirror and a
double-bounce channel-cut Ge(220) crystal monochromator followed by a \SI{0.1}{\milli\metre} divergence slit.
The detector is a 1D LynxEye.
\SI{2.5}{\degree} Soller slits have been used on both the primary and secondary beam
paths.

The macroscopic magnetic properties have been measured within a Quantum Design MPMS XL 5
SQUID magnetometer in the RSO sample transport ($\mu_0 H_\text{max} = \pm\SI{5}{\tesla}$) and
also in a Quantum Design PPMS system ($\mu_0 H_\text{max} = \pm\SI{14}{\tesla}$) with a Vibrating Sample
Magnetometer (VSM) insert. The temperature-dependent magneto-transport
have been probed in the same setup.
The Curie temperature has been measured using a SQUID oven insert.
Point contact Andreev reflection (PCAR) measurements\cite{Soulen1998,Kurt-PRB-2011} have
been performed within the PPMS system and also in a home-built setup using a Nb superconducting tip. Further details on the experimental setup and
data analysis routine can be found elsewhere\cite{Borisov2016b}.

X-ray absorption
spectroscopy (XAS) on the Mn and Fe \ce{L_{2,3}} edges has been carried out at the
European Synchrotron Radiation Facility (ESRF) on the ID32 beam-line\cite{Kummer2016} with both circular left and right
polarisations in order to measure the X-ray magnetic circular dichroism (XMCD). These
measurements have all been carried out at room temperature and in the longitudinal
configuration, \emph{i.e.} with the applied
magnetic field $\mu_0 H= \SI{9}{\tesla}$ collinear to the wave propagation vector
$\vec{k}$. All measurements have been performed in both positive and negative applied magnetic fields.

M\"{o}ssbauer spectroscopy was performed in conversion, at room temperature, using a WissEl
(MA-260) electromagnetic Doppler drive system, a $^{57}$Co(Rh) gamma source, of actual
activity $\sim\SI{40}{\milli\curie}$ and He(\SI{5}{\percent} methane)-gas phase proportional counter, operated at a
fixed pressure of \SI{1.50}{\bar}, and a flow rate of approximately 30 sccm. Canberra amplification and
discrimination electronics were used in conjunction with an in-house developed multi-parameter analyzer
capable of recording simultaneously the Doppler velocity and escape energy for each
detector event, up to
a maximal resolution of $12 \times 12$ bits.
$\alpha$-Fe calibration spectra of 512 channels width, were also acquired to a level
of approximately $10^7$ counts
per channel. Samples were mounted using silver-paint onto aluminium carriers. Custom folding, absorber geometry
modelling, optimal escape energy selection and non-linear least squares regression routines were used for the extraction
of the spectroscopic parameters and their statistical uncertainties. Isomer shifts are referred to the source.

\begin{figure}
\begin{center}
\includegraphics[width=\columnwidth]{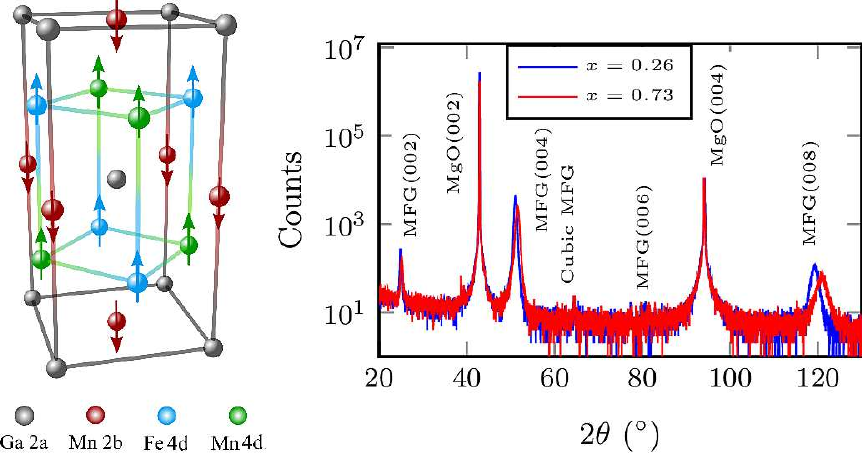}
\caption{
	Left: D0$_{22}$ crystal structure of \ce{Mn2FeGa} assuming perfect ordering. The arrows represent the orientation of the magnetic moments.
	Right: $\theta-2\theta$ XRD scans of two MFG samples with different $x$ values,
confirming the D0$_{22}$ structure, with a fully developed (001) texture.}
\label{fig:structure_xrd}
\end{center}
\end{figure}

\begin{figure*}
\begin{center}
\includegraphics[width=\textwidth]{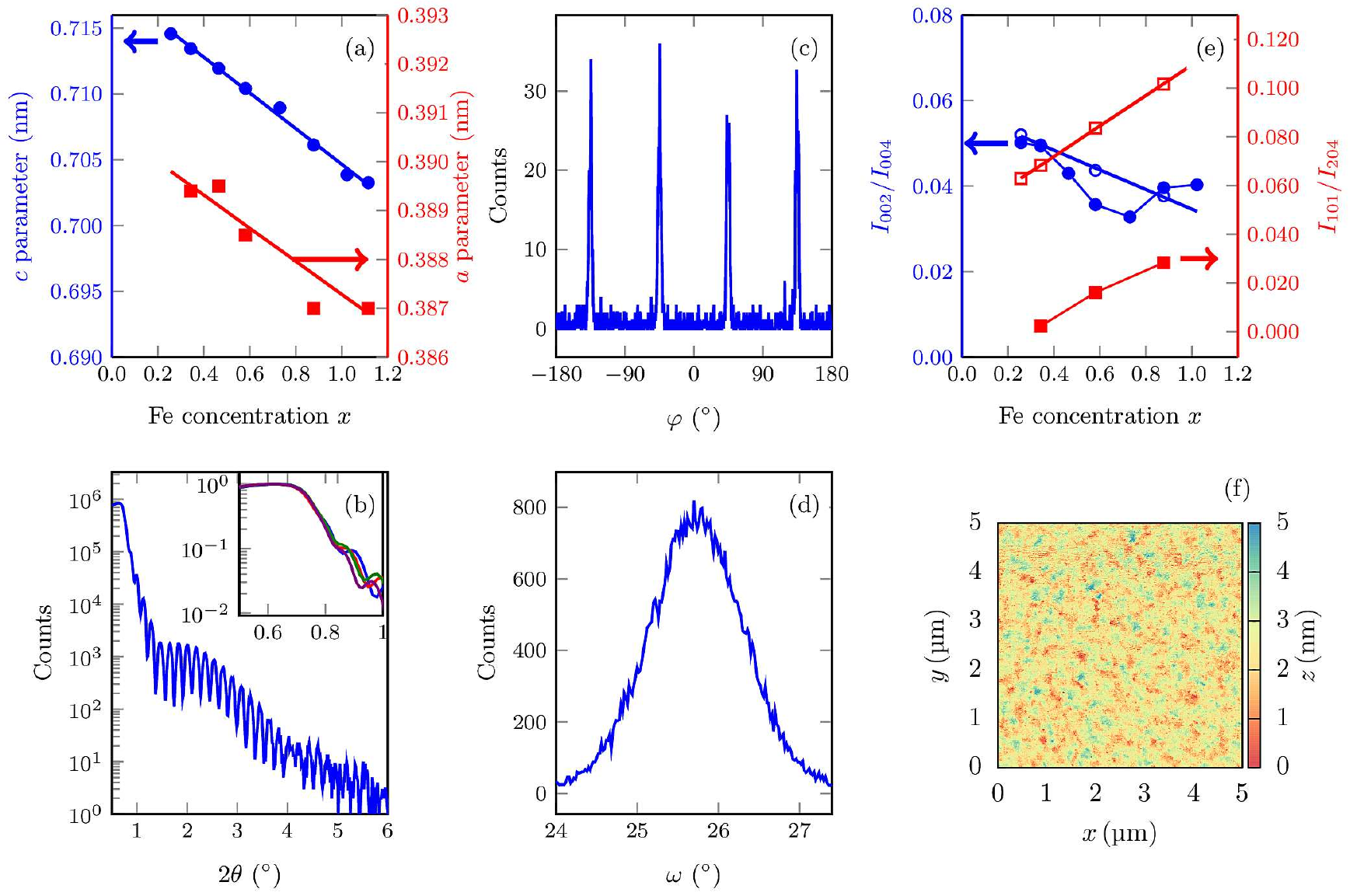}
\caption{(a) $c$ and $a$ lattice parameters as a function of Fe concentration $x$. Solid lines
	are linear regressions.
(b) XRR scan of a \ce{Mn2Fe_{0.58}Ga} thin film. The inset shows the normalised XRR scans of four MFG samples with $x$ ranging from 0.26 to 1.02. The critical angle which is closely related to the
X-ray density is independent of $x$.
(c) $\varphi$ XRD scan of MFG on the MFG(116) peak.
(d) Rocking curve around the MFG(004) peak.
(e) Experimental (filled symbols) and calculated (empty symbols) intensity
ratios for the Bragg peaks as
a function of the Fe concentration $x$. Solid lines are linear regressions of the
calculated intensities.
(f) AFM scan of an MFG sample.}
\label{fig:xrd}
\end{center}
\end{figure*}

\section{Experimental results}\label{sec:structural_magnetic_prop}

\subsection{Structural properties, X-ray diffraction and X-ray reflectivity}\label{subec:XRD}
\ce{Mn2Fe$_x$Ga} is expected to crystallise in the tetragonal D0$_{22}$ structure which belongs to the space
group 139 ($I4/mmm$). The unit cell of a perfectly ordered stoichiometric \ce{Mn2FeGa} is illustrated in the left panel of Fig.\,\ref{fig:structure_xrd}. While the $2a$ Wyckoff position with an octahedral coordination is occupied by \ce{Ga}, \ce{Fe} atoms preferably fill the octahedral $4d$ sites. The $2b$ and the remaining $4d$ sites are filled by \ce{Mn} atoms. If the \ce{Mn} and \ce{Fe} atoms at the $4d$ sites are ordered, as shown in Fig.\,\ref{fig:structure_xrd}, this Wyckoff position can be further separated into $2c$ and $2d$ sublattices. This leads to a structure with lower symmetry, corresponding to the space group 119 ($I-4m2$).

The two, crystallographically inequivalent,
Mn sublattices ($2b$ and $4d$) have been predicted\cite{Wollmann2015} to be antiferromagnetically coupled,
in agreement with what has been experimentally confirmed in closely related compounds such as
\ce{Mn3Ga}\cite{Rode2013},\ce{Mn2Ru$_{x}$Ga}\cite{Betto2015} and \ce{Mn2NiGa}\cite{Paul2014}.
The Fe $4d$ sublattice
has been predicted to be antiferromagnetically coupled to the Mn $2b$ sublattice and ferromagnetically coupled with the Mn $4d$ sublattice\cite{Wollmann2015}.
The overall magnetic structure of MFG can be approximated to that of a collinear ferrimagnet.

The $\theta-2\theta$ symmetric X-ray diffraction data of MFG are shown
in Fig.\,\ref{fig:structure_xrd}. All samples are highly textured with the $c$ axis of the tetragonal unit cell
along the film normal. The three diffraction peaks are indexed, in the $I4/mmm$ space group,
MFG(002), MFG(004), and MFG(008).
We find $a\sim\SI{0.391}{\nano\meter}$ and $c\sim\SI{0.71}{\nano\meter}$,
with both $a$ and $c$ decreasing monotonically with increasing $x$, as shown in Fig.\,\ref{fig:xrd}.
The $\phi$ scan at the MFG(116) reflection (Fig.\,\ref{fig:xrd}(c)) confirms in-plane order.
$\phi=\SI{45}{\degree}$ corresponds to
the $\left[ 110 \right]$ direction of the \ce{MgO} substrate, and we conclude that MFG crystallises in a ``cube-on-cube"
fashion on the \ce{MgO}(100) surface with the in-plane $\ce{MgO} \left[ 110 \right] \parallel \text{MFG} \left[ 110 \right]$.
In Fig.\,\ref{fig:xrd}(d) we show the rocking curve of the MFG(004) peak with a full
width at half maximum of $\sim\SI{1.2}{\degree}$, indicating some degree of mosaicity,
most probably due to the large lattice mismatch ($\sim \SI{7.5}{\percent}$).

The D0$_{22}$ structure  differs from the L1$_0$ by the ordering of the \ce{Ga} atom at ($0.5,0.5,0.5$) fractional coordinates.
The inter-plane ordering of Ga atoms can be estimated by the ordering parameter\cite{Rode2013}:
\begin{eqnarray} \label{eq:S_interplane}
	S_\mathrm{inter-plane}=\sqrt{(I^\textrm{exp}_{002}/I^\textrm{exp}_{004})/(I^\textrm{cal}_{002}/I^\textrm{cal}_{004})}
	\ ,
\end{eqnarray}
while the intra-plane ($2a-2b$) ordering parameter is given by:
\begin{eqnarray} \label{eq:S_intraplane}
S_\mathrm{intra-plane}=\sqrt{(I^\textrm{exp}_{101}/I^\textrm{exp}_{204})/(I^\textrm{cal}_{101}/I^\textrm{cal}_{204})}
\end{eqnarray}
where $I_{hkl}^\text{exp(cal)}$ is the experimental (calculated) intensity of the corresponding Bragg peak.
In Fig.\,\ref{fig:xrd}(e) we show the experimental intensity ratios as a function of Fe
concentration,
together with the expected values in the case of excess Ga atoms in the $2b$ position.
The good agreement between the simulation and the experimental data for $S_\mathrm{inter-plane}$
confirms that Ga is confined to
the $2a-2b$ positions.
The ordering of Ga at the centre of the unit cell is less marked, but still present and it increases with $x$.
Due to the very similar atomic form factors of \ce{Mn} and \ce{Fe}, laboratory X-ray diffractometry is
unable to discern the ordering among these two species.

Previously\cite{Rode2013}, we demonstrated that in manganese-deficient tetragonal \ce{Mn$_{3-x}$Ga} films,
the \ce{Ga} vacancies are negligible and the \ce{Mn} vacancies are distributed over the $2b$ and $4d$ sites,
with a preference for the $2b$ positions. Here, we use the film density, determined from the X-ray reflectometry (XRR), as the key parameter to examine whether this can be applied to MFG.
Using the experimental lattice parameters as inputs, we calculate the expected densities of MFG for $0.26 < x < 1.02$, assuming two extreme cases. First, we assume that deficiencies in Fe systematically lead to the formation of vacancies, \emph{i.e.} similar to the case of \ce{Mn$_{3-x}$Ga}. We find that the MFG density would increase from $\SI{5.9}{\gram\per\centi\metre\cubed}$ for \ce{Mn2Fe$_{0.26}$Ga} to $\SI{7.4}{\gram\per\centi\metre\cubed}$ for \ce{Mn2Fe$_{1.02}$Ga}. Second, we assume instead that the removal of Fe atoms induces a rearrangement of the atomic occupancy such that all atomic sites of the MFG are always fully occupied. In this second case, we find that the density depends only weakly on $x$ and varies from {7.3}
to $\SI{7.4}{\gram\per\centi\metre\cubed}$ for $x = 0.26$ and $x = 1.02$, respectively.

In Fig.\,\ref{fig:xrd}(b) we show a typical grazing-incident XRR scan of a \ce{Mn2Fe_{0.58}Ga} film.
Best fit to the experimental data yields an
X-ray density of \SI{7.6\pm0.1}{\gram\per\centi\metre\cubed} and a low roughness of $\sim\SI{0.5}{\nano\metre}$.
In the inset of Fig.\,\ref{fig:xrd}(b), we show a zoom of the critical angle region of the XRR scans of four MFG films with $x$ ranging from $0.26$ to $1.02$.
The critical angle is the same for all four samples, indicating that their density is almost constant, regardless of the value of $x$.
This observation is in better agreement with the calculated $x$ dependence of the density, assuming that all atomic sites of the MFG are fully occupied.
The slightly higher experimental density may be due to the different sputter yield of \ce{Mn} and \ce{Ga} leading to a ratio lower than $2$,
similar to what found in another
\ce{Mn}-\ce{Ga} based Heusler, doped with \ce{Ru}\cite{Betto2016,Zic2016}.
Excess Ga in the $2b$ positions then explains the lower intra-plane order
parameter of MFG.
Following the structural model outlined above, the site occupancy of some of the samples is reported in
Tab.\,\ref{tab:occupancy}.

\begin{table}
	\caption{Site occupancy as a function of Fe concentration $x$}
  \begin{ruledtabular}
  \begin{tabular}{l c c c c c}
    $x$ & Ga $2a$ & Ga $2b$ & Mn $2b$ & Mn $4d$ & Fe $4d$ \\
    \colrule
	0.26 & 2.00 & 0.45 & 1.55 &  3.36 & 0.63 \\
	0.46 & 2.00 & 0.31 & 1.69 &  2.94 & 1.06 \\
	0.73 & 2.00 & 0.14 & 1.86 &  2.43 & 1.57 \\
	1.00 & 2.00 & 0.00 & 2.00 &  2.00 & 2.00 \\
 \end{tabular}
\end{ruledtabular}
  \label{tab:occupancy}
\end{table}

Thin-film surface/interface quality and roughness are important parameters for integration in spin-electronic device stacks.
We recorded tapping mode AFM images of the MFG surface.
The surface of the film is smooth and free of pinholes
(Fig.\,\ref{fig:xrd}(f)), despite the high crystallinity of MFG and the large lattice mismatch
with the substrate. The extracted RMS
roughness of about \SI{0.6}{\nano\metre} is in good agreement with the XRR fits.
The surface roughness of the films might be further optimised by using lattice-matched substrate
(e.g. \ce{SrTiO3}) or appropriate seed layers (e.g. Pt(001) and IrMn(001))\cite{Kurt-PRB-2011,Jeong2016}.
We note that the low roughness of MFG is in contrast with
D0$_{22}$-\ce{Mn3Ga} films grown directly on \ce{MgO} under similar conditions,
which exhibit discontinuous island-like morphology\cite{Kurt-PRB-2011}. We attribute this to the
Fe incorporation that significantly
improves the surface wetting of Mn-Ga alloys on \ce{MgO} substrate. We further speculate that the change of
the film morphology/wettability between \ce{Mn2Mn$_x$Ga} and \ce{Mn2Fe$_x$Ga} is closely linked to the
different tendencies for creating vacancies or antisites when the two compounds are off-stoichiometric ($x<1$).

\begin{figure}[h!]
	\begin{center}
\includegraphics[width=\columnwidth]{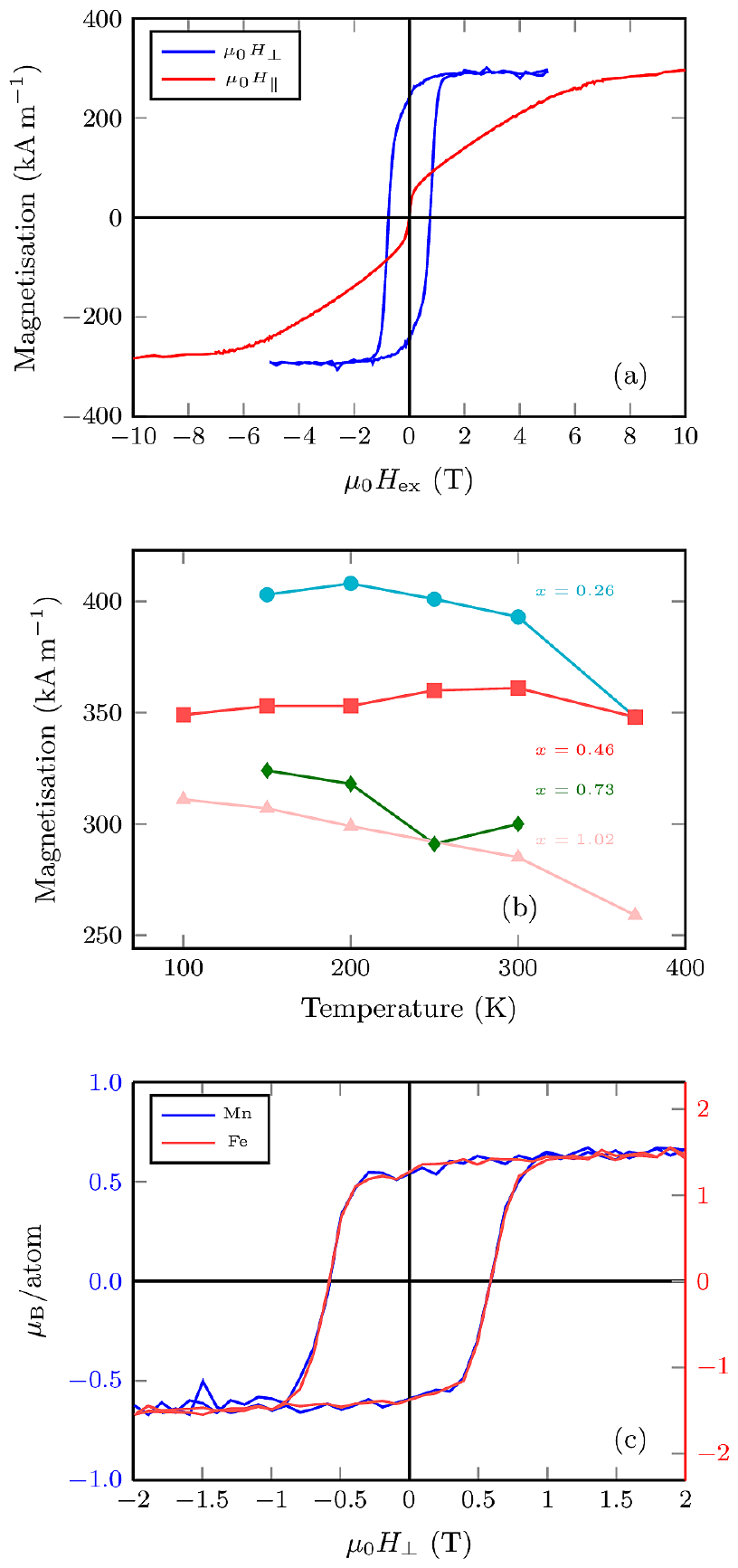}
\caption{(a) Typical out-of-plane and in-plane $M$-$H$ loops of a \ce{Mn2Fe_{0.73}Ga} film.
	(b) Temperature dependence of the magnetisation of MFG films with 4 different Fe concentrations.
	(c) Element-specific hysteresis loops obtained from the XMCD signal of a MFG sample
with $x=0.26$.
}
\label{fig:SQUID}
\end{center}
\end{figure}

\begin{figure}
\centering
\includegraphics[width=\columnwidth]{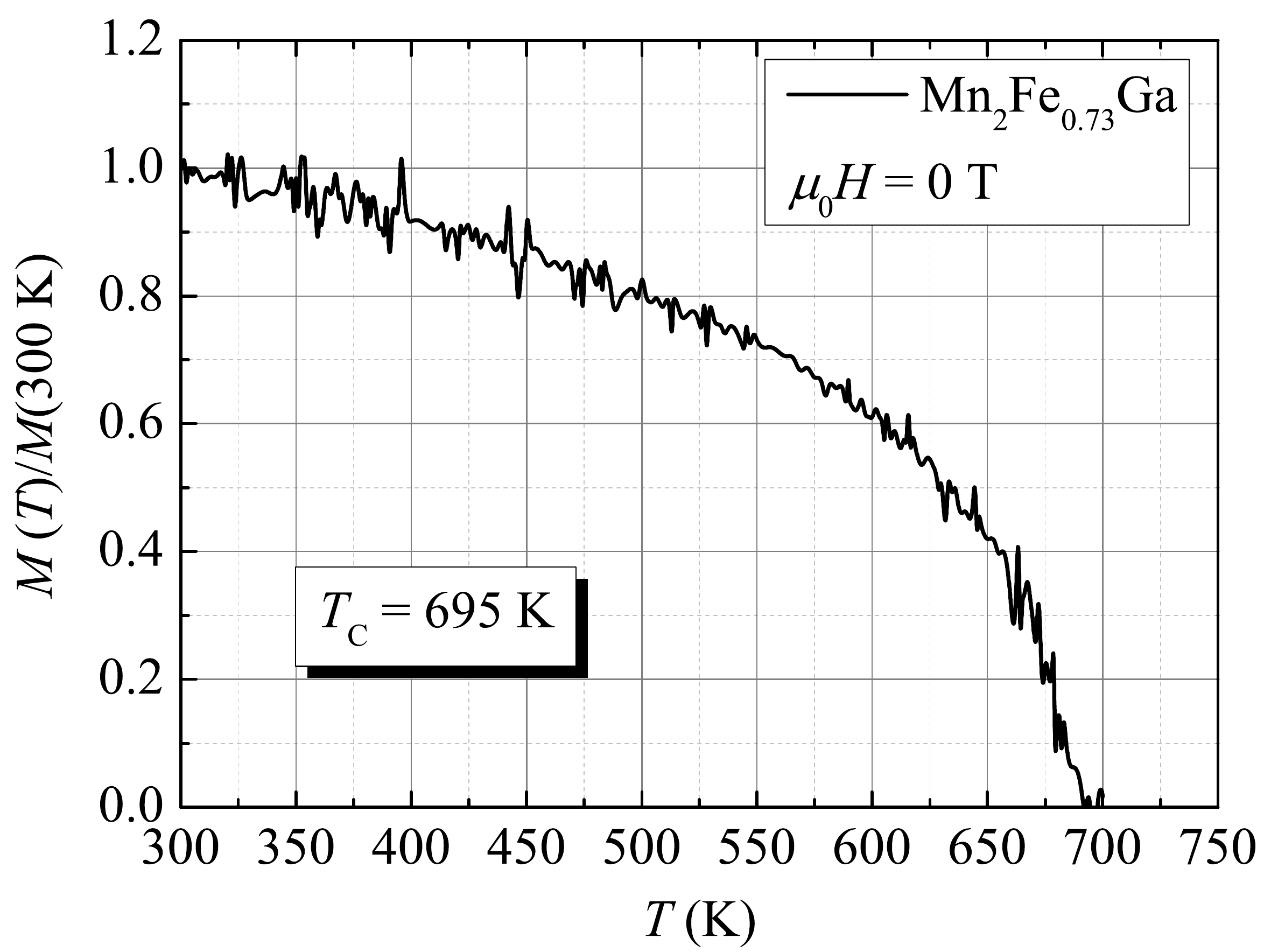}
\caption{SQUID magnetometry of a \ce{Mn2Fe_{0.73}Ga} film in the oven insert. The
	sample was saturated out-of-plane at \SI{300}{\kelvin}, set in remanence and measured
	while heating up. The curve is normalized to its value at
	\SI{300}{\kelvin}. The extracted Curie temperature is \SI{695}{\kelvin}.
}
\label{fig:MFG_oven}
\end{figure}

\begin{figure}[h]
\begin{center}
\includegraphics[width=\columnwidth]{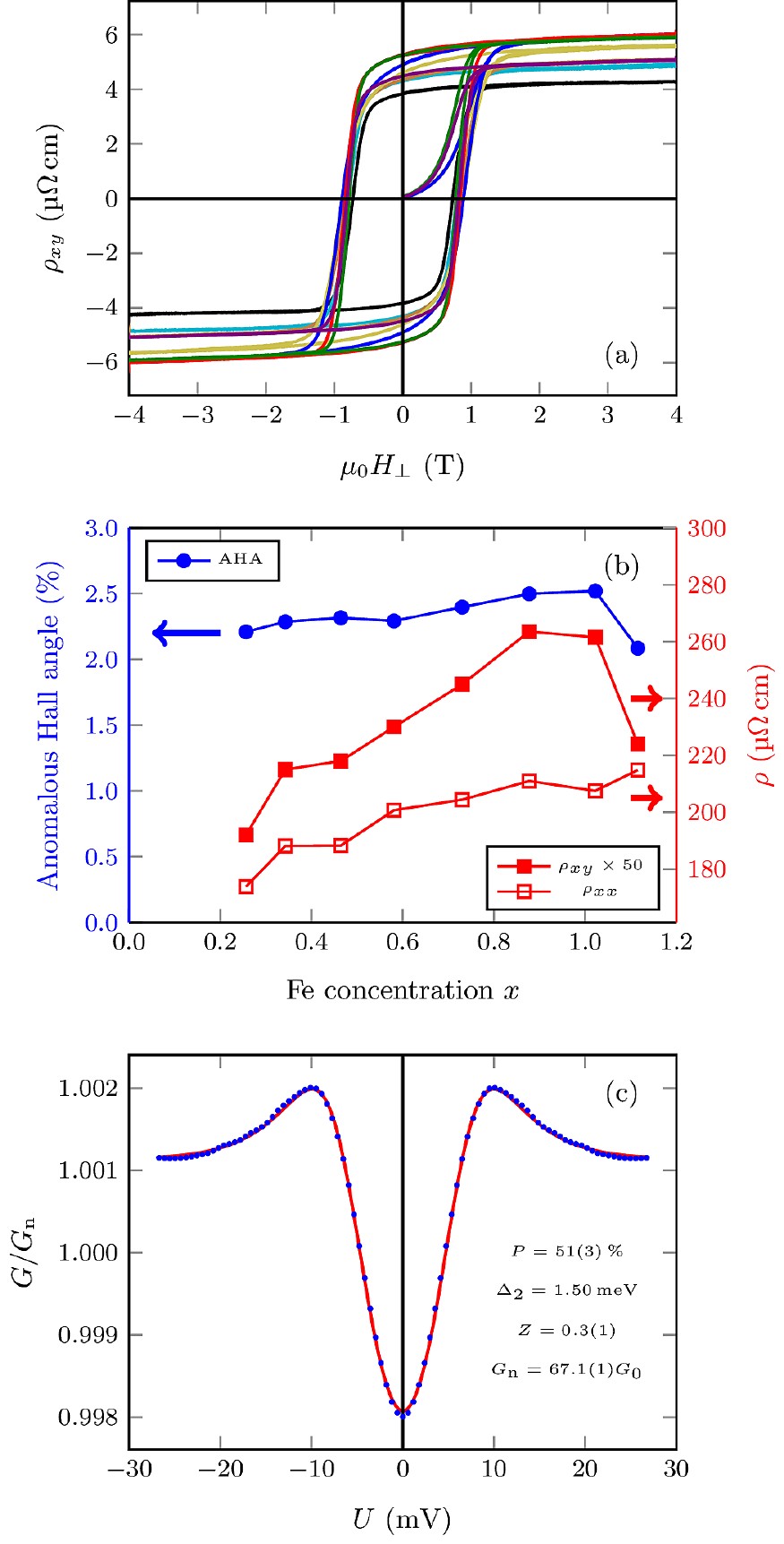}
\caption{(a) Anomalous Hall effect hysteresis loops for samples with different Fe concentration $x$ as a function of the out-of-plane external field $\mu_0H_\perp$.
	(b) Anomalous Hall angle, $\rho_{xy}$ and $\rho_{xx}$ at room temperature as a function of the Fe
	concentration $x$.
	(c) Point contact Andreev reflection spectroscopy of MFG with $x=0.73$. The red line
 is a fit using the parameters listed in the figure. For details see Ref.\cite{Stamenov2012}.
}
\label{fig:transport}
\end{center}
\end{figure}

\subsection{Magnetic and magneto-transport properties}\label{subsec:magnetism}

In Fig.\,\ref{fig:SQUID}(a) we plot the $M$-$H$ loops of a \ce{Mn2Fe_{0.73}Ga} film, measured at \SI{300}{\kelvin} with in-plane $\mu_0H_\parallel$ and out-of-plane $\mu_0H_\perp$ applied fields.
The sample exhibits strong PMA with a low saturation magnetisation $M_\mathrm{s} = \SI{320}{\kilo\ampere\per\metre}$, a coercive field of
\SI{0.85}{\tesla} and a high remanence ratio of $\sim \SI{90}{\percent}$.
From the hard-axis $M$-$H$ loop, we extract an anisotropy field $\mu_0 H_\mathrm{an} = \SI{6.5}{\tesla}$. Using $K_\mathrm{eff} \approx M_\mathrm{s}H_\mathrm{an}/2$ we estimate an effective
anisotropy constant $K_\mathrm{eff}$ that exceeds $\SI{1}{\mega\joule\per\metre\cubed}$. We find that the general shape of the hysteresis loop and the high $K_\mathrm{eff}\geq\SI{1}{\mega\joule\per\metre\cubed}$ are maintained for MFG films with $x$ ranging from $0.26$ to $1.02$.
We should also note that a soft in-plane component is systematically found in the hard $M$-$H_\parallel$ loops of MFG with various Fe concentrations. Previously, similar observations in tetragonal \ce{Mn$_{3-x}$Ga} were attributed to the oscillatory exchange coupling between the first and second nearest \ce{Mn} neighbours\cite{Rode2013}. Our experimental observations suggest that similar situation is likely to happen in MFG, even in the presence of the additional magnetic Fe atoms. This soft component is not observed in anomalous Hall effect loops recorded on the same samples, as Mn in the $2b$ positions contribute only marginally to the Fermi-level density of states\cite{Zic2016}.
In Fig.\,\ref{fig:SQUID}(b), we show the temperature dependence of the saturation magnetisation for samples with different
$x$. The MFG magnetisation at room temperature decreases from $400$ to $\SI{280}{\kilo\ampere\per\metre}$ with increasing Fe doping. The similar temperature dependence of the magnetisation suggests that the Curie temperature of these compounds with varying $x$ remains to be well above the room temperature.

The Curie temperature of a \ce{Mn2Fe_{0.73}Ga} thin film has been directly measured using SQUID magnetometry with an oven insert.
The sample magnetisation was first saturated at \SI{300}{\kelvin} using an out-of-plane \SI{5}{\tesla} field and left
in remanent state. The remanent magnetisation was measured while warming up the sample, as shown in Fig.\,\ref{fig:MFG_oven}.
The extracted Curie temperature is \SI{695}{\kelvin}.
The sample was then cooled to $T = \SI{300}{\kelvin}$ and
a full hysteresis loop was recorded. No changes in
magnetic properties were observed. We therefore conclude that the heat treatment does not affect the magnetic properties of MFG and the extracted Curie temperature is not due to irreversible structural change from D0$_{22}$ (tetragonal-ferrimagnetic) to D0$_{19}$ (hexagonal-antiferromagnetic).


\begin{figure}
\centering
\includegraphics[width=\columnwidth]{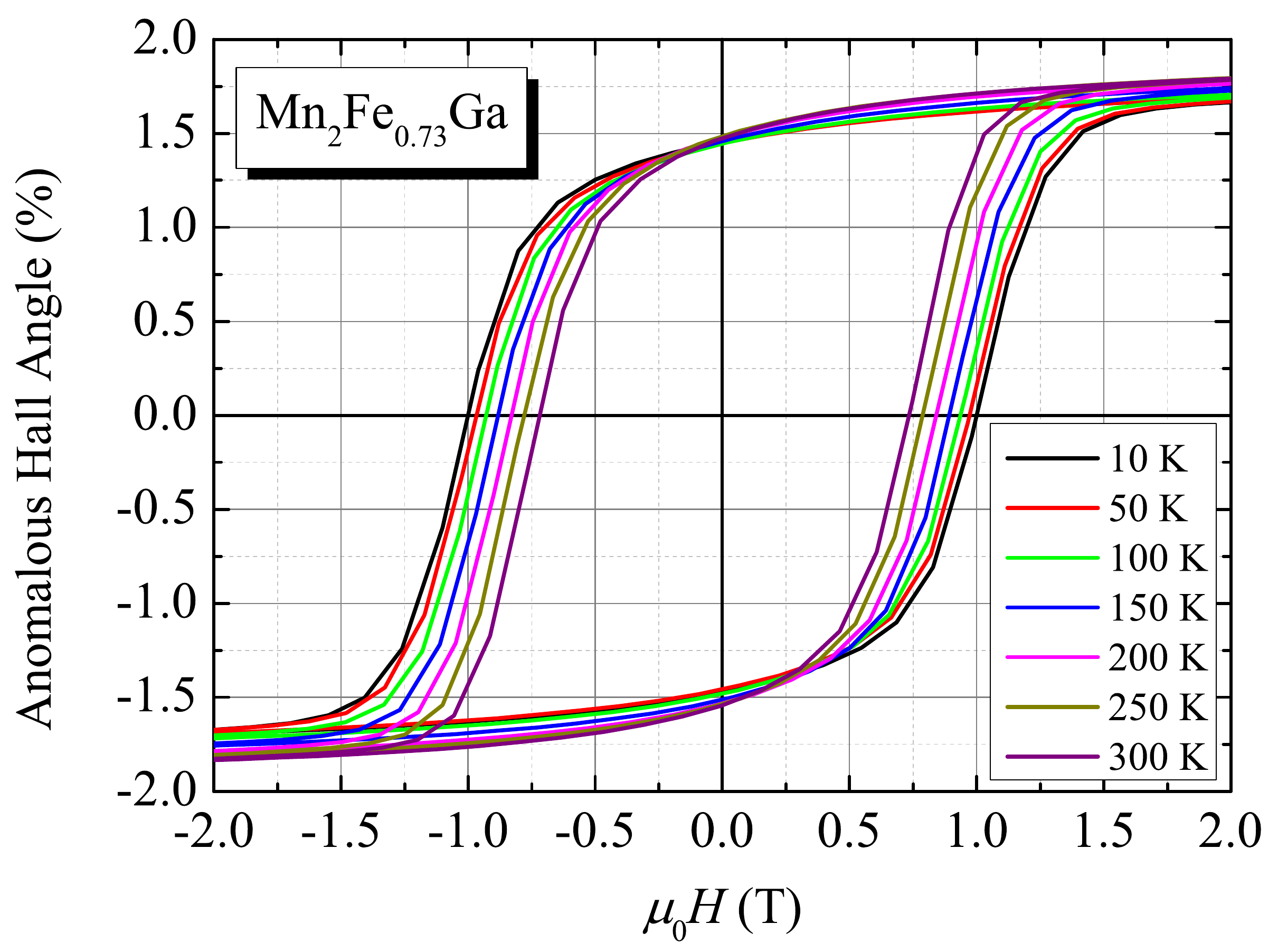}
\caption{Anomalous Hall effect loops for a \ce{Mn2Fe_{0.73}Ga} film at various temperatures ranging from  $10$ to $\SI{300}{\kelvin}$. Each curve was measured up to $\pm \SI{5}{\tesla}$ but
only data within $\pm \SI{2}{\tesla}$ range are shown.
}
\label{fig:EHE_PPMS}
\end{figure}

The magneto-transport of unpatterned MFG films are measured in standard Van der Pauw geometry.
The transverse resistivity $\rho_{xy}$ as a function of $\mu_0H_\perp$ at \SI{300}{\kelvin} for MFG films with $x$ ranging from $0.26$ to $1.12$ are plotted in Fig.\,\ref{fig:transport}(a). The $\rho_{xy}(\mu_0H_\perp)$ hysteresis loops are dominated by the anomalous Hall effect contribution and corroborate the $M-H$ loops obtained from the SQUID magnetometry.
The longitudinal resistivity $\rho_{xx}$, the transverse resistivity at remanence $\rho_{xy}$ and the anomalous Hall angle (AHA), \emph{i.e.} the
$\rho_{xy}/\rho_{xx}$ ratio, of these samples at room temperature are summarised in Fig.\,\ref{fig:transport}(b). $\rho_{xx}$ increases with increasing $x$ which may be attributed to the enhanced scattering causing by the additional \ce{Fe} atoms. $\rho_{xy}$ exhibits similar but steeper dependence with varying $x$, leading to a moderate increase of AHA up to $x=1.02$.
We highlight that AHA is about \SI{2.5}{\percent} at $x=1.02$, which is comparable to
other Mn-Ga based compounds \cite{Thiyagarajah2015,Wu2010} and is an order of
magnitude higher than the AHA normally found in $3d$ transition metal
ferromagnets\cite{Dorleijn1976}.
This value would be even larger if MFG can be further optimised to achieve full remanence.

A linear regression of $\log(\rho_{xy})$ versus $\log(\rho_{xx})$ yields $\rho_{xy} \propto \rho_{xx}^{1.6}$,
which is typically found in weakly localised bad metals\cite{Nagaosa2010}.
The high AHA is indicative of a strong intrinsic contribution (related to Berry phase curvature)
and/or a high spin polarisation at the Fermi level.
Although the tetragonally distorted, Mn-based Heusler alloys have not been predicated to be half-metallic
(\SI{100}{\percent} spin polarisation) unlike their cubic
cousins\cite{Wollmann2014,Wollmann2015},
a pseudo-spin-gap has been theoretically predicted and experimentally
observed\cite{Kurt-PRB-2011}.
To further elucidate the origin of the high AHA, we have measured Point Contact Andreev Reflection (PCAR) \cite{Stamenov2012}
for a MFG film with $x=0.73$.
A typical PCAR spectrum is shown in Fig.\,\ref{fig:transport}(c), for which data analysis following the work of Strijkers \textit{et al}\cite{Strijkers2001,Borisov2016b} yields a high spin polarisation $P=\SI{51}{\percent}$.
This is the highest spin polarisation measured among all
D0$_{22}$ compounds, excluding \ce{Mn3Ga} which, in the tetragonal form, is
metastable and often forms rough and non-continuous films.
The ease of growing continuous and smooth MFG films may prove to be more important
than achieving the highest possible spin polarisation for future spin-electronic device structures.

The temperature dependence of the anomalous Hall effect loops for \ce{Mn2Fe_{0.73}Ga}
is plotted in Fig.\,\ref{fig:EHE_PPMS}. The magnitude of $\rho_{xy}$ is practically temperature independent, with only slight increase of the coercivity at lower temperatures. These observations indicate little Fermi level spin polarisation degradation with increasing temperature and demonstrate the high robustness of the PMA due to the tetragonal D0$_{22}$ structure.
It is worth noting that the magnetism and the transport properties of the tetragonal \ce{Mn2Fe$_x$Ga} differ significantly from those of the near-cubic \ce{Mn2Ru$_x$Ga}, despite the fact that the total valence electron of both systems increases from 17 to 25 when $x$ is varied from 0 to 1. Notably, \ce{Mn2Ru$_x$Ga} possesses a highly tunable magnetic compensation point where both the magnetisation and the anomalous Hall effect change sign\cite{Kurt2014,Thiyagarajah2015}. This can be understood by considering the famous 18 and 24 valence electron counting rule which is valid for half-metallic cubic half-Heusler and full Heusler compounds, respectively\cite{Galanakis2002_halfheusler,Galanakis2002_fullheusler}. Here, we find that such unique properties are absent in tetragonal MFG. Our findings are in better agreement with the theoretical prediction that the magnetism in tetragonal Heuslers approximately obeys a $``25.7"$ valence electron counting rule\cite{Wollmann2015}. The model predicts $\SI{-0.7}{\BohrMagneton\per\formulaunit}$ for D0$_{22}$ \ce{Mn2Fe1Ga}, while we obtain $\SI{-1.8}{\BohrMagneton\per\formulaunit}$ at \SI{100}{\kelvin}.


\subsection{X-ray absorption and magnetic circular dichroism}\label{subsec:xmcd}

XMCD is the method of choice for determining the element-specific spin and orbital angular momenta
in complex thin film systems.
We measured XAS and XMCD at the $L_{3,2}$ edges of Mn and Fe
(corresponding to the electronic transition $2p^63d^n \rightarrow 2p^53d^{n+1}$) for four
samples with $x=0.26$, $0.46$, $0.73$ and $1.02$.
The incident beam $\vec k$-vector has been kept parallel to the applied magnetic field
and at an angle $\theta$ with the sample normal, hence
$\theta = \SI{0}{\degree}$ corresponds to normal incidence.
All data are recorded by detection of the total electron yield (TEY).
Saturation effects have not been taken into account since, for the worst case scenario of
\ce{Mn2Fe1Ga}, the ratio between the electron escape depth and the X-ray absorption
length at the maximum incident angle used,
$\lambda_\mathrm{e}/\lambda_\mathrm{x} \cos\theta$, is estimated to be
$\sim\num{0.03}$, resulting in a correction of about
\SI{2}{\percent}\cite{Nakajima1999,Henke1993}.

We first concentrate on the element-specific hysteresis loops of MFG,
measured on the $L_3$ edges of Mn and Fe.
We find that the Fe and the net Mn moments are strongly and ferromagnetically coupled, with
superimposed hysteresis loops (see Fig.\,\ref{fig:SQUID}(c))
for all values of $x$, in agreement with the colinear magnetic mode described above.

\begin{figure}
	\begin{center}
\includegraphics[width=\columnwidth]{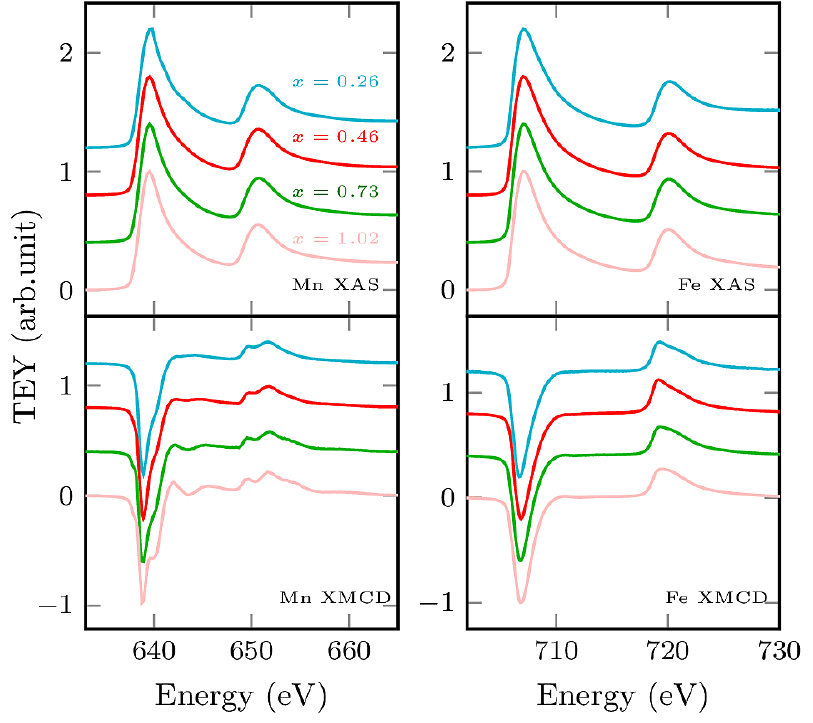}
\caption{XAS and XMCD spectra for Mn and Fe at normal incidence for samples with
different Fe concentration $x$. The XAS spectra are normalised to unity in the post-edge
region. The XMCD spectra are normalised to unity in order to
better appreciate the changes in the shape. All spectra are shifted for clarity.}
\label{fig:spectra}
\end{center}
\end{figure}

In Fig.\,\ref{fig:spectra} we show absorption and dichroism spectra at the Mn and Fe $L$ edges.
The broad and featureless XAS spectral shape indicates that both ions are in a metallic
environment.
Upon increasing $x$ (addition of Fe), the Mn XMCD spectra
become increasingly more structured, indicative of increased electronic localisation.
The Fe XMCD spectra on the other hand remain broad and metallic-like.
It has been shown\cite{Kubler1983} that Mn atoms in the $2b$ positions are more localised
than those in the $4d$ sublattices, which are expected to be of almost completely itinerant nature.
This corraborates our structural model where Fe occupies mainly a subset of the $4d$ positions,
while Mn occupies both $2b$ and $4d$.
As $x$ increases, $4d$ is progressivly emptied of Mn (filled by Fe),
and the ratio of Mn occuping $2b$ increases, resulting in increased Mn localisation.

We now turn to the orbital and spin moments of Fe and Mn as a function of $x$ and $\theta$.
The expectation values of the spin $\langle S \rangle$, orbital $\langle L
\rangle$ and magneti dipole moment  $\langle T \rangle$ were determined
using the magneto-optical sum rules\cite{Carra1993}, from which we infer the spin, orbital and dipolar moments.

From the uniaxial anisotropy constant $K_\mathrm{u}$ and the saturation magnetisation, determined by magnetometry,
we find that the sample magnetisation can be saturated in the direction $\theta$ in an applied field of \SI{9}{\tesla}, available at the beamline.
We can therefore determine the effective spin $m_{S_\mathrm{eff}} = m_S + 7m_T$ and orbital $m_L$ moments,
where $m_S$ and $m_T$ are the spin and magnetic dipole moments, respectively.
At saturation, the spin moment is parallel to the applied field, whereas the orbital and dipole momenta may remain at an angle.
The effective spin moment as a function of incidence angle $\theta$ is:\cite{Koide2004}
\begin{align}
	m_S + 7 m_T^\theta &= m_S + 7 (m_T^\perp\cos^2\theta +
	m_T^\parallel\sin^2\theta)
	\label{eq:angulardep}
\end{align}
where $m_T^\perp$ and $m_T^\parallel$ are the out-of-plane and in-plane components of the magnetic dipole moment.
In particular, we recorded spectra at the `magic' angle $\theta =\SI{54.6}{\degree}$,
where $m_T^\perp + 2 m_T^\parallel = 0$  for point group symmetries higher than $D_{4h}$ and therefore  $m_{S_\mathrm{eff}}=m_S$\cite{Stohr1995}.

\begin{figure}
	\begin{center}
\includegraphics[width=\columnwidth]{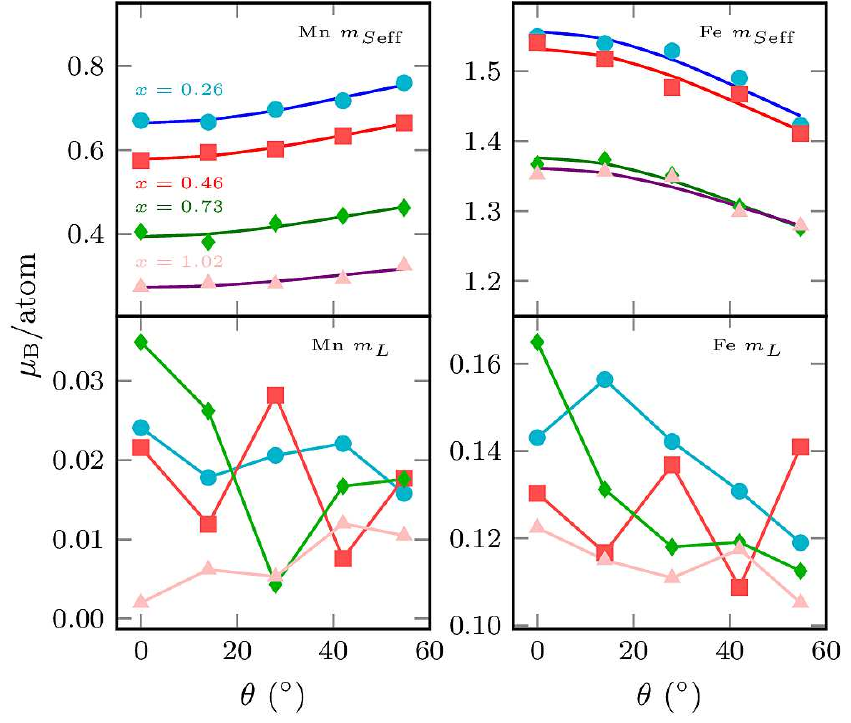}
\caption{Effective spin and orbital moments of the Mn and Fe sublattices as a function of the incidence angle of
	the beam $k$-vector and field direction. Solid lines in the $m_{S_\mathrm{eff}}$
	plots are fits using eq.\,\ref{eq:angulardep}.}
\label{fig:moments}
\end{center}
\end{figure}

The effective spin and orbital momenta determined via the sum rules are shown as a function of incidence angle $\theta$ in Fig.\,\ref{fig:moments}.
The solid lines in the upper panels are fits to the experimental values using Eq.\,\ref{eq:angulardep}.
We find average values of $m_T^\parallel$ of $\sim 0.005$ and $\sim\SI{-0.008}{\BohrMagneton\per atom}$ for Mn
and Fe,
respectively, while the average values of $m_T^\perp$ are $\sim -0.010$ and
$\sim\SI{0.015}{\BohrMagneton\per atom}$.
One can show\cite{Stohr1995} that
for more than half-filled $3d$ shells, a positive (negative) value of $m_T^\parallel$
($m_T^\perp$) corresponds to octahedral (tetrahedral) coordination with tetragonal
distortion, which is consistent with our structural model.
Upon combining the net moments from the sum rules and the Fe content $x$, we have verified that the resulting
net magnetisation for the four samples are in reasonable agreement with those obtained from the magnetometry,
which validates our structural and compositional analyses in Section \ref{subec:XRD}. Ga, as expected, gives rise
to negligible contribution to the magnetisation.

The orbital moments are weak, as expected for ions in a tetragonal environment,
and decrease as $\theta$ increases, except for Mn when $x > 1$.
The anisotropy of the orbital momenta is a direct consequence of magnetocrystalline
anisotropy\cite{Bruno1993},
so that the higher orbital moment is observed along the easy axis of the local environment.
From the angular dependence of $m_L$, we infer that the Fe positions ($4d$),
exhibit easy-$c$-axis anisotropy.
Mn is present in both the $4d$ and $2b$ positions, and for $x > 1$, the anisotropy constant changes sign, going from easy-axis to easy plane.
We have previously found\cite{Rode2013} that for \ce{Mn3Ga} the $4d$ site has perpendicular
anisotropy, while the $2b$ site possesses in-plane anisotropy.
Therefore, when the Mn $2b$ occupation is sufficiently high,
the net magnetocrystalline anisotropy of Mn changes from easy-axis (due to $4d$),
to easy-plane (due to $2b$) in MFG.


Using the results from the sum rules and the magnetometry, we attempt to estimate the average
site-specific magnetic moments and uniaxial anisotropies in MFG, assuming that they remain
constant with increasing Fe dopant occupying $4d$ sites.
The magnetocrystalline anisotropy derives from the second-order correction in the
energy due to the spin-orbit coupling when $m_S$ is rotated from parallel to
perpendicular to the out-of-plane easy axis. It has been evaluated using the Bruno
model\cite{Bruno1989,VanDerLaan1998} and estimating the change in orbital moment from the
equation\cite{Koide2004}:
\begin{align}
	m_L^\theta &= m_L^\perp\cos^2\theta +
	m_L^\parallel\sin^2\theta \ .
	\label{eq:mlangulardep}
\end{align}
The average site-specific $K_\mathrm{u}$ per atom are then renormalised by the site occupancy and the volume of \ce{Mn2FeGa}. Results are reported in
Tab.\,\ref{tab:sitespecificvalues}.


\begin{table}
    \caption{Average site-specific spin moments $m_S$ and uniaxial anisotropies $K_\mathrm{u}$. $K_\mathrm{u}$ in units of \si{\mega\joule\per\metre\cubed} are calculated for \ce{Mn2FeGa}.}
  \begin{ruledtabular}
  \begin{tabular}{l c c c c}
      & Mn $4d$ & Mn $2b$ & Fe $4d$ & Fe $2b$ \\
    \colrule
      \multicolumn{5}{c}{\SI{100}{\percent} Fe $4d$ occupancy} \\
    \colrule
      $m_S$ (\si{\BohrMagneton\per atom}) & 1.5 & 0.9 & 1.4 & -  \\
      $K_\mathrm{u}$ (\si{\milli\electronvolt\per atom})& 0.37 & -0.46 & 0.42 & - \\
      $K_\mathrm{u}$ (\si{\mega\joule\per\metre\cubed})& 1.1 & -1.4 & 1.3 & - \\

    \colrule
      \multicolumn{5}{c}{\SI{80}{\percent} Fe $4d$ occupancy} \\
    \colrule
      $m_S$ (\si{\BohrMagneton\per atom}) & 1.9 & 2.1 & 2.1 & 2.0 \\
      $K_\mathrm{u}$ (\si{\milli\electronvolt\per atom})& 0.50 & -0.86 & 0.53 & -0.11 \\
      $K_\mathrm{u}$ (\si{\mega\joule\per\metre\cubed})& 1.8 & -2.1 & 1.3 & -0.06 \\

 \end{tabular}
\end{ruledtabular}
  \label{tab:sitespecificvalues}
\end{table}

\begin{figure}
\centering
\includegraphics[width=\columnwidth]{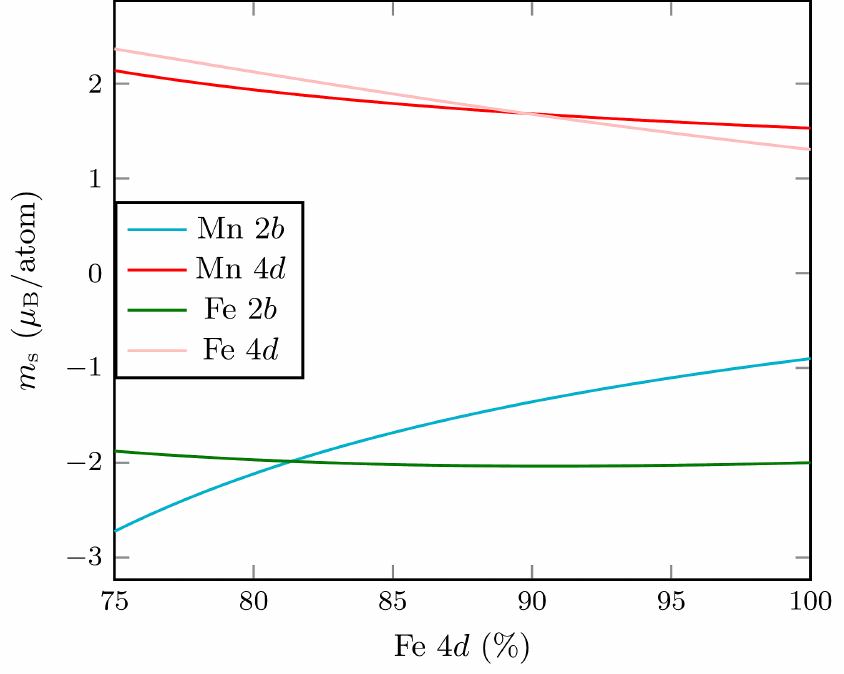}
\caption{Site-specific moment as a function of the occupancy percentage
of Fe atoms in the $4d$ site.}
\label{fig:sitespecificvalues}
\end{figure}

Although adequate in describing the macroscopic properties of the MFG system, this simple model, that assumes a perfectly
ordered crystal, leads to a higher Mn $4d$ moment than that of the Mn $2b$, in contrast with earlier reports\cite{Rode2013} and with the fact that the $3d$ shell of Mn in $2b$ sites is more localised than that of Mn in $4d$ sublattices.
We believe that this is due to the presence of a small amount of Fe $2b$ antisites.
In order to elucidate the consequences due to possible Fe disorder, we show in Fig.\,\ref{fig:sitespecificvalues} the site-specific magnetic moments
assuming a certain percentage of Fe atoms occupying the $2b$ site and coupled antiferromagnetically with their $4d$
counterparts. We find that the Mn $2b$ moment increases more rapidly with increasing Fe disorder and eventually surpasses
the Mn $4d$ moment when about $\SI{20}{\percent}$ of Fe are in $2b$ sites.
In addition, the site-specific Mn moments increases towards the expected values of $\sim
\SI{2}{\BohrMagneton}$\cite{Rode2013}. This degree of Fe site-disorder may explain our M\"{o}ssbauer results in the following section.

\begin{figure}
\centering
\includegraphics[width=\columnwidth]{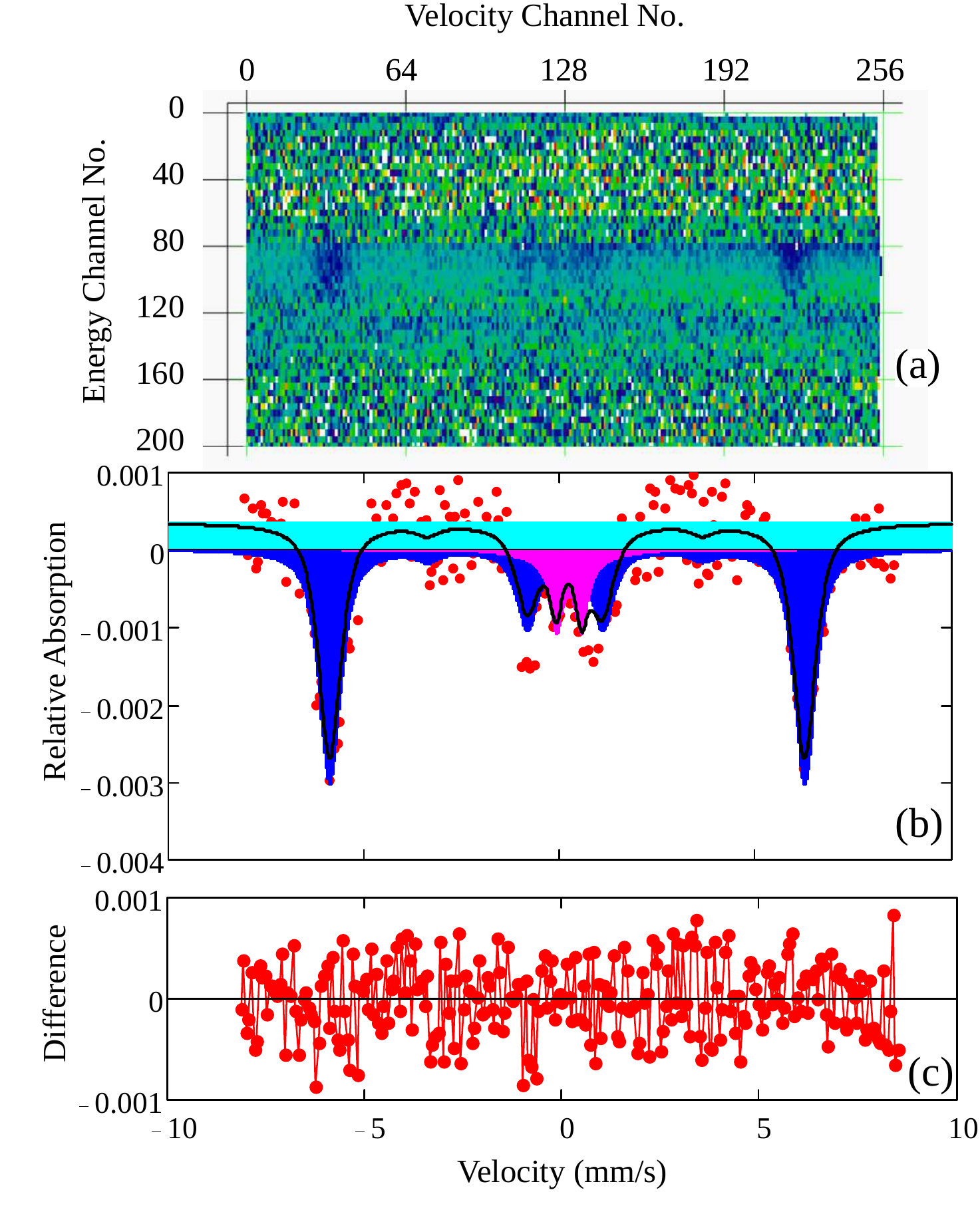}
\caption{(a): Pre-processed two-dimensional M\"{o}ssbauer data.
	(b) Optimal velocity spectrum and two-component fit.
(c) Spectrum-fit difference.}
\label{fig:mossbauer}
\end{figure}

\subsection{M\"{o}ssbauer spectroscopy}\label{subsec:mossbauer}

\begin{figure}
\centering
\includegraphics[width=\columnwidth]{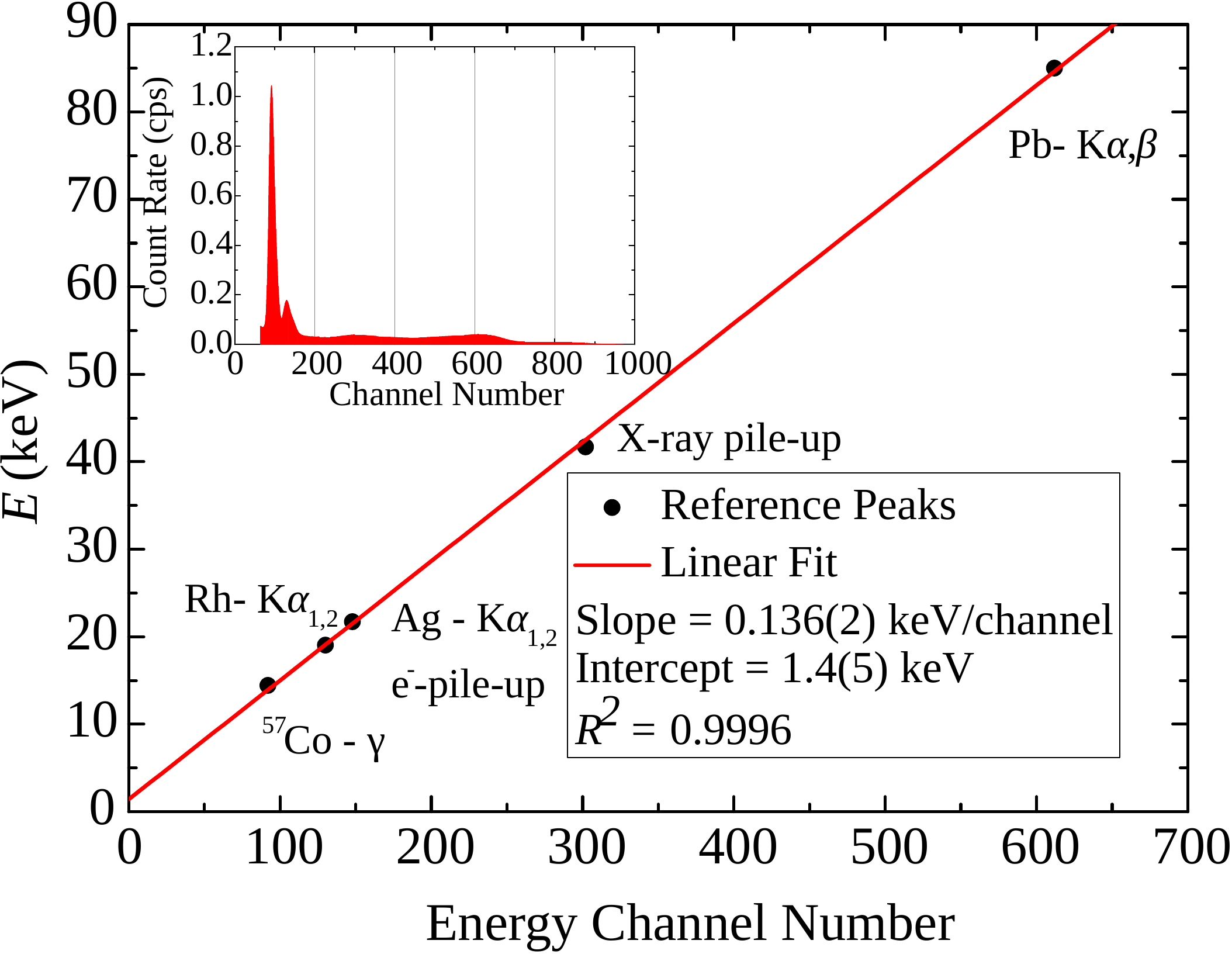}
\caption{Assignment of characteristic X-ray lines for M\"{o}ssbauer and
calibration. Inset: integral energy spectrum.}
\label{fig:mossbauercalibration}
\end{figure}

In order to verify the microscopic magnetic order of the system, M\"{o}ssbauer spectra have been acquired
in the conversion electron geometry on the \ce{Mn2Fe_{0.75}Ga} sample. The multi-parameter analyzer allows for
the optimal software discrimination and weighting of what is otherwise a rather weak absorption from the
sub-percent natural Fe$^{57}$ content. The pre-processed two-dimensional data are shown in
Fig.\,\ref{fig:mossbauer} (a). There is
a single well-defined absorption contrast corresponding to the energy of the incoming gamma rays (centred
around energy channel 90). The clear sextet corresponds to magnetically ordered iron oriented predominantly
out-of-plane. In order to confirm the assignment of the observed signal, the integral energy spectrum, as
represented on the inset of Fig.\,\ref{fig:mossbauercalibration}, is interpreted in terms of the various gamma and characteristic X-ray
lines present. The assignment and calibration are visualized in
Fig.\,\ref{fig:mossbauercalibration}.

The optimal velocity spectrum is shown on Fig.\,\ref{fig:mossbauer} (b), together with a two-component fit to it. The two
components are attributed to ordered out-of-plane metallic iron in a slightly distorted local environment
(effective hyperfine field of \SI{37.7(1)}{\tesla}) and paramagnetic (or in-plane magnetised with a low ordering temperature, in close to
Fe$^{3+}$ configuration). Approximately \SI{85}{\percent} of the iron content is ordered at room temperature, which agrees well
with the measured, by SQUID magnetometry, Curie temperature of \SI{695}{\kelvin} and the
shape of the $M(T)$ curve. The detailed
fitting parameters are listed in Tab.\,\ref{tab:mossbauer}, with their statistical errors
included in brackets.
In details, the parameters are: $\Gamma=$ Lorentzian FWHM, $\sigma=$ Gaussian standard
deviation,
$\mu=$ weighting factor in the pseudo-Voigt lineshape ($\mu \mathrm{Lorentz} + (1-\mu)
\mathrm{Gauss})$), $\delta=$ isomer shift, $\Delta/6=$ electric
quadrupole moment, $B_\mathrm{hf}=$ hyperfine field, $\Theta=$ average angle between the
surface normal (the gamma beam direction) and the local magnetisation.
The \SI{40}{\nano\metre} thick
sample did sustain discharge damage in the course of the first two weeks of data acquisition. No magnetic
M\"{o}ssbauer signature was resolvable in three subsequent one-month-long data runs at
both high and low detector pressure of the same sample.

\begin{table*}
	\caption{Fitting parameters for the M\"{o}ssbauer spectrum}
  \begin{ruledtabular}
  \begin{tabular}{l c c c c c c c c}
	  Model & Area (\si{\percent}) & $\Gamma$ (\si{\milli\metre\per\second}) & $\sigma$
	  (\si{\milli\metre\per\second})  & $\mu$ & $\delta$ (\si{\milli\metre\per\second}) &
	  $\Delta/6$ (\si{\milli\metre\per\second}) & $B_\mathrm{hf}$ (\si{\tesla}) &
	  $\Theta$ (\si{\radian}) \\
    \colrule
	Ferro. sextet & 84.9 & 0.6(2) & 0.3(2) & 0.8 & 0.15(3) & 0.01(1) & 37.7(1) & 0.2(1) \\
	Para. doublet & 15.1 & 0.4(1) & 1(2) & 0.8 & 0.22(5) & 0.11(2) & 0 & - \\
 \end{tabular}
\end{ruledtabular}
  \label{tab:mossbauer}
\end{table*}

\section{Conclusions} \label{sec:conclusions}
In summary, the structural, magnetic and magneto-transport properties of ferrimagnetic \ce{Mn2Fe_xGa} films have
been experimentally investigated.
This ternary compound crystallises in the tetragonally-distorted D0$_{22}$ structure, which results in high
perpendicular anisotropy ($K_\mathrm{u} \geq \SI{1.0}{\mega\joule\per\metre\cubed}$) with a high
in-plane anisotropy field exceeding \SI{6}{\tesla}.
The ferrimagnetic spin structure leads to low net magnetisation at room temperature, which is tunable from $400$ to
\SI{280}{\kilo\ampere\per\metre} with increasing Fe content. The thin films are smooth and pinhole-free, with
a RMS roughness of $\sim \SI{0.6}{\nano\metre}$.
The anomalous Hall angle as high as $\sim \SI{2.5}{\percent}$ has been interpreted as a sign
of high spin polarisation. This has been confirmed by PCAR spectroscopy, which shows an
appreciable \SI{51}{\percent} transport spin polarisation on the same sample.
Using X-ray absorption spectroscopies, the
magnitude and coupling of the spin and orbital magnetic moments for each element have been
determined.
The element-specific characterisation of the moments and the angular dependence allow us
to propose a model for the evolution of the magnetic structure and the site occupancy
with $x$. This model with $\sim \SI{20}{\percent}$ of Fe antisites in $2b$ sublattices is corroborated by the M\"{o}ssbauer spectra.

In ferrimagnetic materials, the precession of the magnetisation presents two
characteristics resonance modes.
The lower frequency mode, also called the ferromagnetic mode, occurs when the
magnetic moments of the two antiferromagnetically-coupled sublattices precess
together as a single, lower amplitude, moment. Its resonance frequency is\cite{Awari2016} $f_- = \gamma/2\pi
(\mu_0 H_\mathrm{an} - \mu_0 M_\mathrm{s}) \sim \SI{150}{\giga\hertz}$, where
$\gamma$ is the gyromagnetic ratio.
The higher frequency mode, or ferrimagnetic mode, is related to the exchange field
and the site-specific moments and anisotropies of the sublattices.
The exchange field can be approximated from the Curie
temperature employing a simple
mean-field model of ferrimagnetism and averaged nearest-neighbours interactions.
Using an exchange field of $\sim\SI{250}{\tesla}$, we estimate the resonance frequency of the ferrimagnetic mode $f_+$ to be above \SI{10}{\tera\hertz}\cite{Wangsness1953}. In this
case, we considered a sample of \ce{Mn2Fe1Ga} and averaged the properties of Mn and
Fe in the $4d$ sublattice.
Interestingly, there seems to be no vacancies in the
crystal structure for any value of $x$, contrarily to what has been found in
\ce{Mn_{3-x}Ga}\cite{Rode2013}.

In summary, Fe-doped Mn-Ga films exhibit the unique combination of outstanding magnetic and transport properties,
in addition to low surface roughness. This novel material with perpendicular anisotropy is an exceptional candidate for
spin-transfer-torque devices, such as memories and oscillators,
with high thermal stability down to \SI{10}{\nano\metre} and ferromagnetic resonance
frequency of about \SI{150}{\giga\hertz}.

\begin{acknowledgments}
D.B., Y.-C.L., K.R. and J.M.D.C. acknowledge the funding by the SFI through AMBER and grant 13/ERC/12561.
K.B. and P.S. acknowledge financial support from Science Foundation Ireland (SFI) within SSPP (11/SIRG/I2130).
\end{acknowledgments}


 \newcommand{\noop}[1]{}

\end{document}